\documentclass[12pt]{article}

\usepackage{graphics}
\usepackage[utf8x]{inputenc}
\usepackage[T2A]{fontenc}
\usepackage[english]{babel}
\usepackage{textcomp}
\usepackage{mathtools}
\usepackage{dcolumn}
\usepackage{epsfig}
\usepackage{epstopdf}
\usepackage{bm}
\usepackage[left=1cm,right=1cm,top=2cm,bottom=2cm]{geometry}
\usepackage[usenames]{color}
\usepackage{colortbl}
\usepackage{mathtools}

\usepackage{amsmath}

\DeclareMathOperator{\arctanh}{Arth}

\begin{document}
\title{Tidal properties of D-dimensional Tangherlini black holes}
	\author{V.P. Vandeev, A.N. Semenova,}
\maketitle
\begin{center}
	{\it Petersburg Nuclear Physics Institute of National Research Centre ``Kurchatov Institute'',\\ Gatchina, 188300, Russia}
\end{center}
\abstract{This paper investigates tidal forces in multidimensional spherically symmetric spacetimes. We consider geodesic deviation equation in Schwarzschild--Tangherlini metric and its electrically charged analog. It was shown that for radial geodesics these equations can be solved explicitly as quadratures in spaces of any dimension. In the case of five, six and seven dimensional spaces, these solutions can be represented in terms of elliptic integrals. For spacetimes of higher dimension, we find the asymptotics of the solution. It was found that in the physical singularity vicinity tidal stretch along the radial direction is the stronger the greater the dimension of space. Whereas the tidal compression in transverse to radial directions, starting from a certain dimension, does not change in the main order. Also in the case of non-radial geodesics, the presence of black hole electric charge does not affect the force of transverse compression in the leading order. For non-radial geodesics with non-zero angular momentum, the local properties of solutions of geodesic deviation equations in the vicinity of a singularity are studied.

\maketitle
\section{\label{Int}Introduction}
General relativity is the theory which describes gravitational phenomena and large-scale structure of spacetime. One of the most interesting objects in the GR is black hole. Different black holes have a huge number of various interesting properties, but in this paper we will focus our attention on the tidal properties of static spherically symmetric black holes. The first work on tidal effects in Schwarzschild spacetime \cite{MS} was paper \cite{SSSS}. These ideas were developed in \cite{SSagain}. In the work \cite{TDE} the influence of tidal forces on the destruction of stars was considered. More fine effects associated with the presence of angular momentum of a test body moving in the gravitational field of a Schwarzschild black hole were investigated in \cite{NRtf}. The case of electrically charged Reissner--Nordstr{\"{o}}m black holes \cite{MR}, \cite{MN} was considered in \cite{TFrn}.Tidal force of black holes in the presence of a homogeneous cosmological constant \cite{MKot} have been studied in \cite{TFkm}.

The tidal properties of black holes surrounded by various matter were also studied: article \cite{TFkbh} is about Kiselev black hole \cite{Kis} surrounded by quintessence, article \cite{TFdbh} is about dirty black holes. The so-called regular black holes, despite the absence of a singularity, have also been studied in the context of tidal forces in \cite{TFrbh}, \cite{chh}, as well as naked singularities in \cite{TFns} and \cite{NN}. The study of geodesic deviation in the theories of modified gravity was carried out in the works \cite{MG} and \cite{EGB}, the first of them considers tidal effects in the theory of holographic massive gravity, the second in theory with the action of Einstein-Gauss-Bonnet. Ref. \cite{ChTF} discusses tidal effects created by dark matter halo. It is important to note that tidal acceleration can be studied not only in metrics of spherical symmetry, but also in axially symmetric spaces. Thus, in \cite{Kerr} tidal forces were studied in the Kerr metric \cite{TFaxkerr}.

There are works in which tidal acceleration is not an object of study, but a means of describing the formation of astrophysical jets: \cite{J1}, \cite{J2} and \cite{J3}. Also one can study affection of tidal forces on a body moving on a finite orbits: \cite{SHef}, \cite{SHefKerr}, \cite{Melk}, \cite{GDL}. These works are devoted to the study of the mechanism of geodesic deviation in various gravitational fields.

 Our main goal is to study the tidal forces of multidimensional Schwarzschild--Tangherlini black holes and their electrically charged generalization \cite{Tangh}. We investigate geodesic deviation equation, which in differential geometry is called the Jacobi equation. We are interested in how the rate of geodesic divergence  depends on the spacetime dimension. Both radial geodesics and geodesics with nonzero angular momentum are considered. The article has the following structure. In Sec.~\ref{GE} we define geodesic equations in $D$-dimensional spherically symmetric spacetime, in Sec.~\ref{GDE} present geodesic deviation equation and find the spectrum of the tidal tensor, which allows us to diagonalize the equation under study by passing to the frame of freely falling body reference. In Sec.~\ref{SGDE} we present solutions of the equation of geodesic deviation at zero angular momentum, if the dimension of space-time allows it. And for non-radial geoedsics with non-zero angular momentum, using the Frobenius method for equations of the Fuch's class, we construct local solutions in the form of power series in the physical singularity vicinity. And in Sec.~\ref{conc} we summarize the results of our study and outline ways of developing the study of tidal forces. The paper also contains two appendices, where we give brief information about the elliptic integrals that were used in the work, Fuchs' equations and the method for constructing their solution.

We use the metric signature $(+,-,...,-)$, greek indices $\alpha,\beta...$ take on the values $0,1,...,D-1$, which correspond hyperspherical coordinates $t,r,\theta_1,...,\theta_{D-2}$ and set the speed of light $c$ and Newtonian gravitational constant $G$ to $1$ throughout this paper.

\section{\label{GE}Geodesics in static spherically symmetric high-dimensional spacetimes}
We consider spacetime with line element of $D$-dimen\-sional static spherically symmetric metric given by
\begin{equation}\label{metric}
ds^2=g_{\mu\nu}dx^{\mu}dx^{\nu}=f(r)dt^{2}-\frac{dr^2}{f(r)}-r^2d\Omega^2_{D-2},
\end{equation}
where $d\Omega^2_{D-2}$ is
\begin{equation}\label{sphere}
d\Omega^2_{D-2}=d\theta^2_1+\sum\limits_{j=2}^{D-2}d\theta^2_j\left(\prod\limits_{k=1}^{j-1}\sin^2\theta_k\right).
\end{equation}
We will define the explicit form of the function $f(r)$ in the following sections. In this spacetime we have the following geodesic equations
\begin{equation}\label{teq}
u^0=\frac{dt}{d\tau}=\frac{E}{f(r)},
\end{equation}
\begin{equation}\label{req}
\left(u^1\right)^2=\left(\frac{dr}{d\tau}\right)^2=E^2-f(r)\left(\delta_1+\frac{L^2}{r^2}\right),
\end{equation}
\begin{equation}\label{theq}
u^j=\frac{d\theta^j}{d\tau}=0,\;j=1,...,D-3,
\end{equation}
\begin{equation}\label{pheq}
u^{D-2}=\frac{d\theta^{D-2}}{d\tau}=\frac{L}{r^2},
\end{equation}
where $\tau$ is affine parameter along the geodesic, $E$ and $L$ are freely moving test body energy and angular momentum. And it should be noted that dynamics of the angular variables $\theta^j$ with $j=1,...,D-3$ are trivial because in a spherically symmetric space movement occurs in one plane determined by the set of equalities $\theta_1=\theta_2=...=\theta_{D-3}=\frac{\pi}{2}$. Azimuth variable dynamic $\theta^{D-2}$ is nontrivial because there is angular momentum $L$. The expression set (\ref{teq})-(\ref{pheq}) forms a unit covariant $D$-velocity vector $u^{\mu}$ tangent to the geodesic.

\section{\label{GDE}Geodesic deviation equation}
 Below we consider geodesic deviation equation. As it is well known \cite{GR}, the equation for the geodesic deviation vector $\tilde{\xi}^{\mu}$ is given by
\begin{equation}\label{deq}
\frac{D^2\tilde{\xi}^{\mu}}{d\tau^2}=R^{\mu}_{.\nu\alpha\beta}u^{\nu}u^{\alpha}\tilde{\xi}^{\beta},
\end{equation}
where $\frac{D^2}{d\tau^2}$ is covariant derivative along the geodesic curve, $R^{\mu}_{.\nu\alpha\beta}$ is Riemann curvature tensor of spacetime and $u^{\nu}$ is the unit vector of $D$-velocity tangent to the geodesic. $\tilde{\xi}^{\mu}$ connects two points on close geodesics which correspond to the same value of the affine parameter $\tau$.

The nonzero Riemann tensor components calculated by the metric (\ref{metric}) are
\begin{equation}
\begin{split}
&R^{0}_{.101}=-\frac{f''}{2f},\\
&R^0_{.202}=-\frac{rf'}{2},\\
&R^0_{.j0j}=-\frac{rf'}{2}\prod\limits_{k=1}^{j-2}\sin^2\theta_k,\\
&R^1_{.212}=-\frac{rf'}{2},\\
&R^1_{.j1j}=-\frac{rf'}{2}\prod\limits_{k=1}^{j-2}\sin^2\theta_k,\\
&R^i_{.jij}=\left(1-f\right)\prod\limits_{k=1}^{j-2}\sin^2\theta_k,\\
&i=2,...,D-2,\;j=3,...,D-1,
\end{split}
\end{equation}
where prime means derivative with respect to radial variable $r$.

It is seen that on the right-hand side of the Eq.~(\ref{deq}) there is a matrix $P^{\mu}_{.\beta}\equiv R^{\mu}_{.\nu\alpha\beta}u^{\nu}u^{\alpha}$, which is called tidal tensor, have non-zero elements
\begin{equation}
\begin{split}
&P^{0}_{.0}=\frac{\dot{r}^2f''}{2f}+\frac{\chi L^2 \sin^2\theta}{r^2},\\
&P^{0}_{.1}=-\frac{E\dot{r}f''}{2f^2},\;P^{1}_{.0}=\frac{E\dot{r}f''}{2}\\
&P^{1}_{.1}=\frac{\chi L^2 \sin^2\theta}{r^2}-\frac{E^2f''}{2f},\\
&P^{0}_{.D-1}=\frac{\chi E L}{r^2},\;P^{D-1}_{.0}=-\frac{\chi E L \sin^2\theta}{f},\\
&P^{1}_{.D-1}=-\frac{\chi L \dot{r}}{fr^2},\;P^{D-1}_{.1}=-\chi L \dot{r}\sin^2\theta,\\
&P^{j}_{.j}=\frac{(f-1)L^2\sin^2\theta}{r^4}-\chi\omega,\;P^{D-1}_{.D-1}=-\chi\omega,\\
&j=2,...,D-2,
\end{split}
\end{equation}
where
\begin{subequations}
\begin{equation}\label{r1}
\chi=\frac{f'}{2r},
\end{equation}
\begin{equation}\label{r2}
\omega=\delta_1+\frac{L^2}{r^2},
\end{equation}
\begin{equation}\label{r3}
\dot{r}=u^1=\frac{dr}{d\tau}=\sqrt{E^2-f(r)\left(\delta_1+\frac{L^2}{r^2}\right)}.
\end{equation}
\end{subequations}
Therefore, a parallelpropagated tetrad ($D$-dimensional analogue of 4-dimensional tetrads, which we will call simply tetrads below) basis for the free fall frame of reference can be constructed. It has form
\begin{subequations}\label{tetrads}
\begin{equation}\label{tett}
e_{t}^{\mu}=\frac{1}{\sqrt{\delta_1}}\left(\frac{E}{f},\dot{r},0,\frac{L}{r^2}\right),
\end{equation}
\begin{equation}\label{tetr}
e_{r}^{\mu}=\frac{1}{\sqrt{\delta_1}}\left(-\frac{\dot{r}}{f},-E,0,0\right),
\end{equation}
\begin{equation}\label{tetaz}
e_{\theta_j}^{\mu}=\left(0,0,\frac{1}{r},0\right),\;j=1,...,D-3,
\end{equation}
\begin{equation}\label{tetpol}
e_{\theta_{D-2}}^{\mu}=\frac{L}{r\sqrt{\delta_1^2+\frac{\delta_1L^2}{r^2}}}\left(\frac{E}{f},\dot{r},0,\frac{\delta_1+\frac{L^2}{r^2}}{L}\right).\\
\end{equation}
\end{subequations}
These tetrads $e^{\mu}_{\alpha}$ satisfy normalization condition ${e^{\mu}_{\alpha}\:e^{\nu}_{\beta}\:g_{\mu\nu}=\eta_{\alpha\beta}}$ with Minkowski metric $\eta_{\alpha\beta}=diag\left(1,-1,...,-1\right)$.
The geodesic deviation vector $\tilde{\xi}^{\mu}$ can be substituted as
\begin{equation}\label{lt}
    \tilde{\xi}^{\mu}=e^{\mu}_{\nu}\:\xi^{\nu}.
\end{equation}
Thus, the meaningful components of Eq. (\ref{deq}) in this frame of reference are
\begin{equation}\label{radeq}
\ddot{\xi}^r=\left[-\frac{f''}{2}\left(\delta_1+\frac{L^2}{r^2}\right)+\frac{f'}{2r}\frac{L^2}{r^2}\right]\xi^r,
\end{equation}
\begin{equation}\label{azeq}
\ddot{\xi}^a=\left[-\frac{f'}{2r}\left(\delta_1+\frac{L^2}{r^2}\right)+\frac{f-1}{r^2}\frac{L^2}{r^2}\right]\xi^a,
\end{equation}
\begin{equation}\label{poleq}
\ddot{\xi}^{\theta_{D-2}}=-\frac{\delta_1f'}{2r}\xi^{\theta_{D-2}},
\end{equation}
where $a={\theta_1},...,{\theta_{D-3}}$ correspond to polar angular components and temporary component is trivial $\ddot{\xi}^t=0$.
These equations are the diagonal form of Eq.~(\ref{deq}). It is worth noting that the dependence on all angular coordinates have disappeared from Eqs.~(\ref{radeq})--(\ref{poleq}) because there is no dynamics along the direction of the azimuthal angles $\theta^j$ with $j=1,...,D-3$ according to Eq.~(\ref{theq}) and geodesics lie in the equatorial plane $\theta_1=\theta_2=...=\theta_{D-3}=\frac{\pi}{2}$. Also one need to pay attention that without angular momentum $L$ Eq.~(\ref{azeq}) and Eq.~(\ref{poleq}) coincide. The right parts of these equations describe tidal accelerations.
\section{\label{SGDE}Solution of geodesic deviation equation}
Using an Eq.~(\ref{req}) left side of Eqs.~(\ref{radeq})--(\ref{poleq}) can be transformed into
\begin{equation}
\ddot{\xi}^\mu=\left[E^2-f\left(\delta_1+\frac{L^2}{r^2}\right)\right]{\xi^\mu}''-\left[\frac{f'}{2}\left(\delta_1+\frac{L^2}{r^2}\right)-\frac{fL^2}{r^3}\right]{\xi^\mu}',
\end{equation}
where prime means derivative with respect to variable $r$.
\subsection{Radial geodesics}
Radial geodesics are a special class of geodesics. The absence of angular momentum $L$ makes Eqs.~(\ref{radeq})--(\ref{poleq}) easier
\begin{equation}\label{radeqwl}
\left(E^2-\delta_1f\right){\xi^r}''-\frac{\delta_1f'}{2}{\xi^r}'+\frac{\delta_1f''}{2}\xi^r=0,
\end{equation}
\begin{equation}\label{poleqwl}
\left(E^2-\delta_1f\right){\xi^j}''-\frac{\delta_1f'}{2}{\xi^j}'+\frac{\delta_1f'}{2r}\xi^{j}=0,
\end{equation}
where $j=\theta_1,...,\theta_{D-2}$. These equations can be solved through quadratures
\begin{equation}\label{genradsol}
\xi^r=\sqrt{E^2-\delta_1f}\left(A_r+B_r\int\frac{dr}{\left(E^2-\delta_1f\right)^\frac{3}{2}}\right),
\end{equation}
\begin{equation}\label{genangsol}
\xi^j=r\left(C_j+N_j\int\frac{dr}{r^2\sqrt{E^2-\delta_1f}}\right),
\end{equation}
where $A_j,B_j,C_j,N_j$ are integration constants. These solutions have been known for a long time and have been repeatedly used in works \cite{SSSS}, \cite{TFrn}, \cite{TFkm}, \cite{TFkbh}, \cite{Kis}, \cite{MG} and \cite{EGB}. It worth noting lightlike geodesics with $\delta_1=0$ deviate in the same way in space of any dimension
\begin{equation}
\xi^\mu=A+Br
\end{equation}
for all spatial components $\mu=r,\theta_1,...,\theta_{D-2}$.

Previously in a large number of articles expressions (\ref{genradsol}) and (\ref{genangsol}) were applied to 4-dimensional spacetimes. Below in spaces of higher dimensions for timelike geodesics at $\delta_1=1$ we represent them as normal elliptic integrals where it is possible.
\subsubsection{D-dimensional Schwarzschild spacetimes}
Electrically neutral Schwarzschild–Tangherlini metric in the form (\ref{metric}) has function $f(r)$ as
\begin{equation}\label{ShTg}
f(r)=1-\frac{\mu_D}{r^{D-3}},\;\mu_D=\frac{16\pi G_D M}{(D-2)A_{D-2}},
\end{equation}
where $A_{D-2}$ is the surface area of the unit $S^{D-2}$ sphere given as
\begin{equation}\label{SD}
A_{D-2}=\frac{2\pi^{\frac{D-1}{2}}}{\Gamma\left(\frac{D-1}{2}\right)}.
\end{equation}
Expressions (\ref{genradsol}) and (\ref{genangsol}) in spaces of different dimensions are:
\begin{enumerate}
\item For 4-dimensional spacetime
\begin{equation}\label{elsol4r}
\xi^r=A\alpha_4(r)+B\left[3\mu_4+r\psi^2+\frac{3\alpha_4(r)}{\psi}\ln\left(\frac{\alpha_4(r)-\psi}{\alpha_4(r)+\psi}\right)\right],
\end{equation}
\begin{equation}\label{elsol4a}
\xi^a=r\bigg(C+N\alpha_4(r)\bigg),
\end{equation}
where
\begin{equation}
\psi=\sqrt{E^2-1},\;\alpha_4(r)=\sqrt{E^2-1+\frac{\mu_4}{r}}.
\end{equation}
This solution was first obtained in work \cite{SSSS}.
\item For 5-dimensional spacetime solutions are expressed in elementary functions
\begin{equation}\label{elsol5r}
\xi^r=A\alpha_5(r)+Br\left[\alpha^2_5(r)+\frac{\mu_5}{r^2}\right],
\end{equation}
\begin{equation}\label{elsol5a}
\xi^a=r\bigg[C+N\ln\left(\alpha_5(r)+\frac{\sqrt{\mu_5}}{r}\right)\bigg],
\end{equation}
where
\begin{equation}
\alpha_5(r)=\sqrt{E^2-1+\frac{\mu_5}{r^2}}.
\end{equation}
\item  For 6-dimensional spacetime
\begin{equation}\label{elsol6r}
\xi^r=\alpha_6(r)\left[A+BJ_1\left(\frac{\kappa}{r}\right)\right]-B\left(\frac{3\psi^2r}{10\kappa}+\frac{2\kappa^2}{r^2}\right),
\end{equation}
\begin{equation}\label{elsol6a}
\xi^a=r\left[C+NJ_0\left(\frac{\kappa}{r}\right)\right].
\end{equation}
where
\begin{equation}
\kappa=\sqrt[3]{\frac{\mu_6}{4}},\;\psi=\sqrt{E^2-1},\;\alpha_6(r)=\sqrt{E^2-1+\frac{\mu_6}{r^3}},
\end{equation}
and functions $J_0$ and $J_1$ are Weierstrass elliptic integrals, which are defined in App.~\ref{ElInt}, which have $g_2=0$ and $g_3=-\psi^2$.
\item  For 7-dimensional spacetime
\begin{equation}\label{elsol7r}
\xi^r=B\left(\frac{2r}{\omega}+\frac{3\omega^3}{r^3}\right)+\alpha_7(r)A+\alpha_7(r)Be^{\frac{\pi i}{4}}\left[E\left(\frac{\omega e^{\frac{\pi i}{4}} }{r},i\right)-F\left(\frac{\omega e^{\frac{\pi i}{4}} }{r},i\right)\right],
\end{equation}
\begin{equation}\label{elsol7a}
\xi^a=r\left[C+NF\left(\frac{\omega e^{\frac{\pi i}{4}}}{r},i\right)\right],
\end{equation}
where
\begin{equation}
\omega=\sqrt[4]{\frac{\mu_7}{E^2-1}},\;\alpha_7(r)=\sqrt{E^2-1+\frac{\mu_7}{r^4}}.
\end{equation}
And functions $E(x,k)$ and $F(x,k)$ are Legendre elliptical integrals, which defined in App.~\ref{ElInt}, which have argument $x=\omega e^{\frac{\pi i}{4}}/r$ and parameter $k=i$.
\item  For $D$-dimensional spacetime with $D\ge8$ integrals in (\ref{genradsol}) and (\ref{genangsol}) are no longer elliptical. Therefore, below we demonstrate the asymptotic properties of these solutions in the vicinity of a physical singularity and spatial infinity.

At spatial infinity $r\to\infty$ they are
\begin{equation}\label{eq36}
\xi^r=Br+A+\frac{B\sigma_D\left(D-1\right)}{2\left(D-4\right)r^{D-4}}+\frac{A\sigma_D}{2r^{D-3}}-\frac{B\sigma_D^2\left(D-1\right)\left(D-5\right)}{4\left(D-4\right)\left(2D-7\right)r^{2D-7}}-\frac{A\sigma_D^2}{8r^{2D-6}}+O\left(r^{10-3D}\right),
\end{equation}
\begin{equation}\label{eq37}
\xi^a=Cr+N-\frac{N\sigma_D}{\left(2D-4\right)r^{D-3}}+\frac{3N\sigma_D^2}{\left(16D-40\right)r^{2D-6}}+O\left(r^{9-3D}\right),
\end{equation}
where $\sigma_D=\frac{\mu_D}{E^2-1}$. The linear growth of all spatial geodesic deviation vector components at infinity is expected, because Schwarzschild–Tangherlini metric is asymptotically flat.

Near physical singularity $r\to0$ they are
\begin{equation}\label{38}
\xi^r=\frac{A}{r^{\frac{D-3}{2}}}+\frac{Br^{D-2}}{3D-7}+\frac{A\varepsilon_D r^{\frac{D-3}{2}}}{2}-\frac{2(D-2)B\varepsilon_D r^{2D-5}}{(3D-7)(5D-13)}-\frac{A\varepsilon_D^2r^{\frac{3(D-3)}{2}}}{8}+O\left(r^{3D-8}\right),
\end{equation}
\begin{equation}\label{39}
\xi^a=Cr+\frac{Nr^{\frac{D-3}{2}}}{D-5}-\frac{N\varepsilon_Dr^\frac{3(D-3)}{2}}{2\left(3D-11\right)}+\frac{3N\varepsilon_D^2r^\frac{5(D-3)}{2}}{8\left(5D-17\right)}+O\left(r^\frac{7(D-3)}{2}\right),
\end{equation}
where $\varepsilon_D=\frac{E^2-1}{\mu_D}$ and $a=\theta_1,..\theta_{D-2}$. It should be noted that the radial component in the vicinity of $r=0$ increases indefinitely, but the angular components decrease. A large dimension $D$ of spacetime corresponds to a more intensive growth of $\xi^r$.
\end{enumerate}
\subsubsection{D-dimensional Reissner--Nordstr{\"{o}}m spacetimes}
Electrically charged Reissner--Nordstr{\"{o}}m–Tangherlini metric in the form (\ref{metric}) has function $f(r)$ as
\begin{equation}\label{RNTg}
f(r)=1-\frac{\mu_D}{r^{D-3}}+\frac{q_D^2}{r^{2(D-3)}},
\end{equation}
where $q_{D}$ is
\begin{equation}
q_D^2=\frac{8\pi G_D Q^2}{(D-2)(D-3)A_{D-2}},
\end{equation}
and $\mu_D$ and $A_{D-2}$ are defined in Eqs.~(\ref{ShTg}) and (\ref{SD}) respectively.
General expressions (\ref{genradsol}) and (\ref{genangsol}) in spaces of different dimensions take the form:
\begin{enumerate}
\item For 4-dimensional spacetime
\begin{equation}\label{elsolch4r}
\begin{split}
&\xi^r=\zeta_4(r)\left[A-\frac{3B\mu_4}{\left(E^2-1\right)^{\frac{3}{2}}}\arctanh\left(\frac{\left(E^2-1\right)+\frac{\mu_4}{2r}}{\zeta_4(r)\sqrt{E^2-1}}\right)\right]+\\
&+B\left[2r+\frac{\mu_4\left(\mu_4^2\Psi^2+5\right)}{E^2-1}-\frac{2q^2\left(\mu_4^2\Psi^2+2\right)}{r\left(E^2-1\right)}\right],
\end{split}
\end{equation}
\begin{equation}\label{elsolch4a}
\xi^a=r\left\{C+N\arcsin\left(\Psi\left[\mu_4-\frac{2q_4^2}{r}\right]\right)\right\},
\end{equation}
where
\begin{equation}
\begin{split}
&\Psi^2=\left(4q_4^2E^2-4q_4^2+\mu_4^2\right)^{-1},\\
&\zeta_4(r)=\sqrt{E^2-1+\frac{\mu_4}{r}-\frac{q_4^2}{r^2}}.
\end{split}
\end{equation}
These solutions were first obtained and well studied in work \cite{TFrn}.

\item For 5-dimensional spacetime
\begin{equation}\label{elsolch5r}
\begin{split}
&\xi^r=A\zeta_5(r)-\frac{B\zeta_5^2(r)r}{\sqrt{E^2-1}}+\frac{B\beta\zeta_5(r)}{b}\times\\
&\times\left[F\left(x,k\right)-E\left(x,k\right)+\frac{\Pi(x,k,1)-b^8\Pi\left(x,k,ib^{-2}\right)}{1+b^4}\right],
\end{split}
\end{equation}
\begin{equation}\label{elsolch5a}
\xi^a=r\left(C+NF\left(x,k\right)\right),
\end{equation}
where
\begin{equation}
\begin{split}
&x=\frac{\beta}{br},\;k=ib^2,\;\beta=\sqrt[4]{\frac{q^2_5}{E^2-1}},\\
&b^2=\frac{\mu_5}{2q_5\sqrt{E^2-1}}+\sqrt{1+\frac{\mu_5^2}{4q_5^2\left(E^2-1\right)}},\\
&\zeta_5(r)=\sqrt{E^2-1+\frac{\mu_5}{r^2}-\frac{q_5^2}{r^4}},
\end{split}
\end{equation}
and $E(x,k)$, $F(x,k)$  and $\Pi(x,k,c)$ also defined in App.~\ref{ElInt}.
\item For $D$-dimensional spacetime with $D\ge6$ integrals in (\ref{genradsol}) and (\ref{genangsol}) are not elliptical. Therefore, below we demonstrate the asymptotic properties of these solutions in the vicinity of a physical singularity and spatial infinity.

    At spatial infinity $r\to\infty$ they are
\begin{multline}\label{48}
\xi^r=Br+A+\frac{B\sigma_D\left(D-1\right)}{2\left(D-4\right)r^{D-4}}+\frac{A\sigma_D}{2r^{D-3}}-\\
-\frac{B\left[\sigma_D^2\left(D-1\right)\left(D-5\right)+4w_D\left(D-2\right)\left(D-4\right)\right]}{4\left(D-4\right)\left(2D-7\right)r^{2D-7}}
-\frac{A\left(\sigma_D^2+4w_D\right)}{8r^{2D-6}}+O\left(r^{10-3D}\right),
\end{multline}
\begin{equation}\label{49}
\xi^a=Cr+N-\frac{N\sigma_D}{\left(2D-4\right)r^{D-3}}+\frac{N\left(3\sigma^2_D+4w_D\right)}{\left(16D-40\right)r^{2D-6}}+O\left(r^{9-3D}\right),
\end{equation}
where $\sigma_D=\frac{\mu_D}{E^2-1}$ and $w_D=\frac{q^2_D}{E^2-1}$. The presence of black hole electric charge, as expected, did not change spatial components behavior at infinity, because Reissner--Nordstr{\"{o}}m--Tangherlini metric is also asymptotically flat. Therefore, the character of (\ref{48}) and (\ref{49}) coincides with the electrically neutral case (\ref{eq36}) and (\ref{eq37}) respectively.

Near physical singularity $r\to0$ they are
\begin{equation}\label{50}
\xi^r=\frac{A}{r^{D-3}}-\frac{A\varepsilon_D}{2h_D}-\frac{A\left(\varepsilon^2_D+4h_D\right)r^{D-3}}{8h^2_D}-\frac{A\left(\varepsilon^3_D+4\varepsilon_Dh_D\right)r^{2D-6}}{16h^3_D}+\frac{Br^{2D-5}}{3D-8}+O\left(r^{3D-8}\right),
\end{equation}
\begin{equation}\label{51}
\xi^a=Cr+\frac{Nr^{D-3}}{D-4}+\frac{N\varepsilon_Dr^{2D-6}}{\left(4D-14\right)h_D}+\frac{N\left(3\varepsilon^2_D+4h_D\right)r^{3D-9}}{\left(24D-80\right)h^2_D}+O\left(r^{4D-12}\right),
\end{equation}
where $\varepsilon_D=\frac{E^2-1}{\mu_D}$, $h_D=\frac{E^2-1}{q^2_D}$ and $a=\theta_1,..,\theta_{D-2}$. Compared with (\ref{38}) and (\ref{39}) the case of a charged black hole shows us that the presence of a charge does not change angular components behavior (\ref{51}), but enhances the tidal stretch along the radial direction (\ref{50}).

\end{enumerate}
\subsection{Nonradial geodesics}\label{EQfucl}
We now turn to the study of Eqs.~(\ref{radeq})-(\ref{poleq}) in the presence of non-zero angular momentum $L$. In this case these equations cannot be solved in quadratures, so we will look for local solutions in the form of generalized power series using the Frobenius method (briefly about it in App.~\ref{FEQ}).

Equations (\ref{radeq})-(\ref{poleq}) for timelike geodesics $\delta_1=1$ can be rewritten as
\begin{equation}\label{FuEQ}
{\xi^i}''+P(r){\xi^i}'+Q_j(r){\xi^i}=0.
\end{equation}
where
\begin{subequations}\label{C}
\begin{equation}
P(r)=-\frac{\frac{f'}{2}\left(1+\frac{L^2}{r^2}\right)-\frac{fL^2}{r^3}}{E^2-f\left(1+\frac{L^2}{r^2}\right)},
\end{equation}
\begin{equation}
Q_r(r)=\frac{\frac{f''}{2}\left(1+\frac{L^2}{r^2}\right)-\frac{f'}{2r}\frac{L^2}{r^2}}{E^2-f\left(1+\frac{L^2}{r^2}\right)},
\end{equation}
\begin{equation}
Q_a(r)=\frac{\frac{f'}{2r}\left(1+\frac{L^2}{r^2}\right)-\frac{f-1}{r^2}\frac{L^2}{r^2}}{E^2-f\left(1+\frac{L^2}{r^2}\right)},
\end{equation}
\begin{equation}
Q_{\theta_{D-2}}(r)=\frac{\frac{f'}{2r}}{E^2-f\left(1+\frac{L^2}{r^2}\right)},
\end{equation}
\end{subequations}
and indecies $i,j$ take all spatial values $r,a,\theta_{D-2}$. It is easy to see that the coefficients $P(r)$ and $Q_j(r)$ of the Eq.~(\ref{FuEQ}) are rational functions. We are interested in the deviation geodesic vector $\xi^i$ behavior in the vicinity physical singularity $r=0$. We will explore the equation for all spatial components in both metrics (\ref{ShTg}) and (\ref{RNTg}).
\subsubsection{D-dimensional Schwarzschild spacetimes}
Coefficients (\ref{C}) of differential Eq.~(\ref{FuEQ}) in multidimensional Schwarzschild spacetime are
\begin{subequations}\label{csh}
\begin{equation}\label{gencoef}
P(r)=-\frac{1}{2r}\cdot\frac{\mu_DL^2(D-1)+\mu_D(D-3)r^2-2L^2r^{D-3}}{\mu_DL^2+\mu_D r^2-L^2r^{D-3}+(E^2-1)r^{D-1}},
\end{equation}
\begin{equation}\label{rcoef}
Q_r(r)=-\frac{\mu_D(D-3)}{2r^2}\cdot\frac{L^2(D-1)+(D-2)r^2}{\mu_DL^2+\mu_Dr^2-L^2r^{D-3}+(E^2-1)r^{D-1}},
\end{equation}
\begin{equation}\label{pcoef}
Q_a(r)=\frac{\mu_D}{2r^2}\cdot\frac{L^2(D-1)+(D-3)r^2}{\mu_DL^2+\mu_Dr^2-L^2r^{D-3}+(E^2-1)r^{D-1}},
\end{equation}
\begin{equation}\label{azcoef}
Q_{\theta_{D-2}}(r)=\frac{\mu_D(D-3)}{2\big[\mu_DL^2+\mu_Dr^2-L^2r^{D-3}+(E^2-1)r^{D-1}\big]},
\end{equation}
\end{subequations}
where $a=\theta_1,..,\theta_{D-3}$.

Consider all components of Eq.~(\ref{FuEQ}):
\begin{itemize}
\item
The radial equation uses coefficients (\ref{gencoef}) and (\ref{rcoef})
\begin{equation}
{\xi^r}''+P(r){\xi^r}'+Q_r(r){\xi^r}=0.
\end{equation}
To investigate the properties of a given equation solution in the vicinity of a physical singularity $r=0$ using the Frobenius method one need to expand the coefficients in a Laurent series
\begin{subequations}
\begin{equation}\label{gen1}
P(r)=-\frac{D-1}{2r}+O(1),
\end{equation}
\begin{equation}
Q_r(r)=-\frac{(D-1)(D-3)}{2r^2}+O(r^{-1}).
\end{equation}
\end{subequations}
Therefore, the equation for the leading powers of linearly independent solutions takes the form
\begin{equation}
\zeta(\zeta-1)-\left(\frac{D-1}{2}\right)\zeta-\frac{(D-1)(D-3)}{2}=0.
\end{equation}
Its solutions are $\zeta_1=D-1$, $\zeta_2=\frac{3-D}{2}$. The difference of these roots is not integer, so two local linearly independent solutions are represented as
\begin{subequations}
\begin{equation}\label{58a}
\xi_1^r(r)=r^{D-1}\sum_{k=0}^{\infty}c_kr^{k},
\end{equation}
\begin{equation}\label{58b}
\xi_2^r(r)=r^{\frac{3-D}{2}}\sum_{k=0}^{\infty}d_kr^{k},
\end{equation}
\end{subequations}
where the set of coefficients $c_k$ and $d_k$ can be found using the initial data.

Around $r=0$ for any $D$ the first solution is negligible compared to the second, because it has negative powers of the radial variable $r$. Thus, the dimension of spacetime affects the deviation of geodesics along the radial direction. It can be seen that with $L\ne0$ (\ref{58b}) in leading order coincides with (\ref{38}).
\item
The polar equations use coefficients (\ref{gencoef}) and (\ref{pcoef})
\begin{equation}
{\xi^a}''+P(r){\xi^a}'+Q_a(r){\xi^a}=0,
\end{equation}
where $a=\theta_1,...\theta_{D-3}$. Similarly, we use the Laurent series expansion of the coefficient $P(r)$ founded in (\ref{gen1}) and Laurent series for the coefficient $Q_a(r)$
\begin{equation}
Q_a(r)=\frac{D-1}{2r^2}+O(r^{-1}).
\end{equation}
Therefore, the equation for the leading powers of linearly independent solutions takes the form
\begin{equation}
\zeta(\zeta-1)-\left(\frac{D-1}{2}\right)\zeta+\frac{D-1}{2}=0.
\end{equation}
Its solutions are $\zeta_1=\frac{D-1}{2}$, $\zeta_2=1$. The difference of these roots is not integer, so two local linearly independent solutions can be represented as
\begin{subequations}
\begin{equation}\label{62a}
\xi_1^a(r)=r^{\frac{D-1}{2}}\sum_{k=0}^{\infty}c_kr^{k},
\end{equation}
\begin{equation}\label{62b}
\xi_2^a(r)=\sum_{k=0}^{\infty}d_kr^{k+1},
\end{equation}
\end{subequations}
where $a=\theta_1,...\theta_{D-3}$ and the set of coefficients $c_k$ and $d_k$ can be found using the initial data.

In the vicinity of a physical singularity $r=0$ the leading solution is $\xi_2^a(r)$ does not depend on the dimension of spacetime $D$. Linear growth of (\ref{62b}) is similar to (\ref{49}). This means that in the leading order the behavior of the polar components $\xi^a,\;a=\theta_1,..\theta_{D-3}$ does not depend on the presence of angular momentum.
\item
The azimuthal equation uses coefficients (\ref{gencoef}) and (\ref{azcoef})
\begin{equation}
{\xi^{\theta_{D-2}}}''+P(r){\xi^{\theta_{D-2}}}'+Q_{\theta_{D-2}}(r){\xi^{\theta_{D-2}}}=0.
\end{equation}
Similarly, we use the Laurent series expansion of the coefficient $P(r)$ founded in (\ref{gen1}) and Laurent series for the coefficient $Q_{\theta_{D-2}}(r)$
\begin{equation}
Q_{\theta_{D-2}}(r)=\frac{D-3}{2L^2}+O(r).
\end{equation}
It can be seen that the coefficient at $r^{-2}$ is absent.

Therefore, the equation for the leading powers of linearly independent solutions takes the form
\begin{equation}
\zeta(\zeta-1)-\left(\frac{D-1}{2}\right)\zeta=0.
\end{equation}
Its solutions are $\zeta_1=\frac{D+1}{2}$, $\zeta_2=0$. The difference of these roots is not integer, so two local linearly independent solutions are represented as
\begin{subequations}
\begin{equation}\label{66a}
\xi_1^{\theta_{D-2}}(r)=r^{\frac{D+1}{2}}\sum_{k=0}^{\infty}c_kr^{k},
\end{equation}
\begin{equation}\label{66b}
\xi_2^{\theta_{D-2}}(r)=\sum_{k=0}^{\infty}d_kr^{k},
\end{equation}
\end{subequations}
where the set of coefficients $c_k$ and $d_k$ can be found using the initial data.

Around $r=0$ for any $D$ the first solution is negligible compared to the second, therefore the azimuth component of the geodesic deviation vector is constant. And this component behavior differs from (\ref{49}) at $a=\theta_{D-2}$. This means that the azimuthal component $\xi^{\theta_{D-2}}$ depends on $L$. And along this direction, a freely falling body does not experience tidal compression because $\xi^{\theta_{D-2}}\propto const$.
\end{itemize}
A common property of the geodesic deviation vector transverse components $\xi^a\propto r,\;a=\theta_1,..,\theta_{D-3}$ and $\xi^{\theta_{D-2}}\propto const$ is that its behavior near a physical singularity $r=0$ is independent on the spacetime dimension, in contrast to longitudinal component $\xi^r\propto\frac{1}{\sqrt{r^{D-3}}}$.
\subsubsection{D-dimensional Reissner--Nordstr{\"{o}}m spacetimes}
Coefficients (\ref{C}) of differential Eq.~(\ref{FuEQ}) in multidimensional Reissner--Nordstr{\"{o}}m spacetime are
\begin{small}
\begin{subequations}\label{crn}
\begin{equation}\label{gencoefrn}
P(r)=-\frac{D-3}{2r}\cdot\frac{\mu_Dr^{D-3}\left[r^2+\frac{(D-1)L^2}{D-3}\right]-2q^2_D\left[r^2+\frac{(D-2)L^2}{D-3}\right]}{(E^2-1)r^{2D-4}-L^2r^{2D-6}+\left(\mu_Dr^{D-3}-q_D^2\right)\left(r^2+L^2\right)},
\end{equation}
\begin{equation}\label{rcoefrn}
Q_r(r)=-\frac{D-3}{2r^2}\cdot\frac{\mu_Dr^{D-3}\left[(D-2)r^2+L^2(D-1)\right]-4q_D^2\left[(D-\frac{5}{2})r^2+(D-2)L^2\right]}{(E^2-1)r^{2D-4}-L^2r^{2D-6}+\left(\mu_Dr^{D-3}-q_D^2\right)\left(r^2+L^2\right)},
\end{equation}
\begin{equation}\label{pcoefrn}
Q_a(r)=\frac{D-3}{2r^2}\cdot\frac{\mu_Dr^{D-3}\left[r^2+\frac{(D-1)L^2}{D-3}\right]-2q_D^2\left[r^2+\frac{(D-2)L^2}{D-3}\right]}{(E^2-1)r^{2D-4}-L^2r^{2D-6}+\left(\mu_Dr^{D-3}-q_D^2\right)\left(r^2+L^2\right)},
\end{equation}
\begin{equation}\label{azcoefrn}
Q_{\theta_{D-2}}(r)=\frac{D-3}{2}\cdot\frac{\mu_{D}r^{D-3}-2q_{D}^2}{(E^2-1)r^{2D-4}-L^2r^{2D-6}+\left(\mu_Dr^{D-3}-q_D^2\right)\left(r^2+L^2\right)},
\end{equation}
\end{subequations}
\end{small}
where $a=\theta_1,..,\theta_{D-3}$.

Consider all components of Eq.~(\ref{FuEQ}):
\begin{itemize}
\item
The radial equation uses coefficients (\ref{gencoefrn}) and (\ref{rcoefrn}).
\begin{equation}
{\xi^r}''+P(r){\xi^r}'+Q_r(r){\xi^r}=0.
\end{equation}
To investigate the properties of a given equation solution in the vicinity of a physical singularity $r=0$ using the Frobenius method we expand the coefficients in a Laurent series
\begin{subequations}
\begin{equation}\label{genrpol}
P(r)=-\frac{D-2}{r}+O(1),
\end{equation}
\begin{equation}
Q_r(r)=-\frac{2(D-3)(D-2)}{r^2}+O(r^{-1}).
\end{equation}
\end{subequations}
Therefore, the equation for the leading powers of linearly independent solutions takes the form
\begin{equation}
\zeta(\zeta-1)-\zeta(D-2)-2(D-3)(D-2)=0.
\end{equation}
Its solutions are $\zeta_1=2D-4$, $\zeta_2=3-D$. The difference of these roots is integer, so the larger of the roots determine the leading power of the first solution, a logarithmic term may appear in the second linearly independent solution
\begin{subequations}
\begin{equation}\label{71a}
\xi_1^r=r^{2D-4}\sum_{k=0}^{\infty}c_kr^{k},
\end{equation}
\begin{equation}\label{71b}
\xi_2^r=r^{3-D}\sum_{k=0}^{\infty}d_kr^{k}+A\xi_1^r(r)\ln(r),
\end{equation}
where the set of coefficients $c_k$, $d_k$ and $A$ can be found using the initial data. It should be noted that for any $D\ge4$ in the vicinity of the singularity $2D-4$ is positive, and hence the product $r^{2D-4}\ln(r)$ tends to zero, so the leading term is proportional $r^{3-D}$ that goes to infinity for any dimension. This result is similar to (\ref{50}).
\end{subequations}
\item
The polar equations use coefficients (\ref{gencoefrn}) and (\ref{pcoefrn}).
\begin{equation}
{\xi^a}''+P(r){\xi^a}'+Q_a(r){\xi^a}=0,
\end{equation}
where $a=\theta_1,...\theta_{D-3}$.
Similarly, we use the Laurent series expansion of the coefficient $P(r)$ founded in (\ref{genrpol}) and Laurent series for the coefficient $Q_a(r)$
\begin{equation}
Q_a(r)=\frac{D-2}{r^2}+O(r^{-1}).
\end{equation}
Therefore, the equation for the leading powers of linearly independent solutions takes the form
\begin{equation}
\zeta(\zeta-1)-\zeta(D-2)+(D-2)=0.
\end{equation}
Its solutions are $\zeta_1=D-2$, $\zeta_2=1$. The difference of these roots is integer, so the larger of the roots will determine the leading power of the first solution, a logarithmic term may appear in the second linearly independent solution
\begin{subequations}
\begin{equation}\label{75a}
\xi_1^a=r^{D-2}\sum_{k=0}^{\infty}c_kr^{k},
\end{equation}
\begin{equation}\label{75b}
\xi_2^a=\sum_{k=0}^{\infty}d_kr^{k+1}+A\xi_1^r(r)\ln(r),
\end{equation}
\end{subequations}
where $a=\theta_1,...\theta_{D-3}$, the set of coefficients $c_k$, $d_k$ and $A$ can be found using the initial data. It should be noted that for any $D\ge4$ in the vicinity of the singularity $D-2$ is bigger then 1, and hence the product $r^{D-2}\ln(r)$ tends to zero faster then $r$, so the leading term is proportional to $r$. This linear growth of polar geodesic deviation vector components coincide in the leading order with Eq.~(\ref{51}) at $a=\theta_1,..,\theta_{D-3}$.
\item
The azimuthal equation uses coefficients (\ref{gencoefrn}) and (\ref{azcoefrn})
\begin{equation}
{\xi^{\theta_{D-2}}}''+P(r){\xi^{\theta_{D-2}}}'+Q_{\theta_{D-2}}(r){\xi^{\theta_{D-2}}}=0.
\end{equation}
Similarly, we use the Laurent series expansion of the coefficient $P(r)$ found in (\ref{genrpol}) and Laurent series for the coefficient $Q_{\theta_{D-2}}(r)$
\begin{equation}
Q_{\theta_{D-2}}(r)=\frac{D-3}{L^2}+O(r).
\end{equation}
It can be seen that the coefficient at $r^{-2}$ is absent.
Therefore, the equation for the leading powers of linearly independent solutions takes the form
\begin{equation}
\zeta(\zeta-1)-\zeta(D-2)=0.
\end{equation}
Its solutions are $\zeta_1=D-1$, $\zeta_2=0$. The difference of these roots is integer, so the larger of the roots determine the leading power of the first solution, a logarithmic term may appear in the second linearly independent solution
\begin{subequations}
\begin{equation}\label{79a}
\xi_1^{\theta_{D-2}}(r)=r^{D-1}\sum_{k=0}^{\infty}c_kr^{k},
\end{equation}
\begin{equation}\label{79b}
\xi_2^{\theta_{D-2}}(r)=\sum_{k=0}^{\infty}d_kr^{k}+A\xi_1^r(r)\ln(r),
\end{equation}
\end{subequations}
where the set of coefficients $c_k$, $d_k$ and $A$ can be found using the initial data. It should be noted that for any $D\ge4$ in the vicinity of the singularity $D-1$ is bigger then zero, and hence the product $r^{D-1}\ln(r)$ is infinitesimal at $r\to0$, so the leading term is proportional to constant. As in the electrically neutral case, the behavior of the azimuthal component $\xi^{\theta_{D-2}}$ changed with the appearance of the angular momentum from (\ref{51}) to (\ref{79b}).
\end{itemize}
A common property of the geodesic deviation vector transverse components is that its behavior near a physical singularity is independent on the spacetime dimension $D$ $\xi^a\propto r,\;a=\theta_1,..,\theta_{D-3}$ and $\xi^{\theta_{D-2}}\propto const$, in contrast to longitudinal component $\xi^r\propto\frac{1}{r^{D-3}}$.
\section{\label{conc}Conclusion}
In this paper, we consider the geodesic deviation equation (\ref{deq}) in the $D$-dimensional spherically symmetric Tangherlini spacetime. First of all, we show that a parallelpropagated tetrad can be constructed. It describes the transition to a reference frame where the equation under study acquires a digonal form (\ref{tetrads}). This made it possible to obtain expressions (\ref{radeq})-(\ref{poleq}) for all spatial components of the tidal force. In Sec.~(\ref{SGDE}), we solve the geodesic deviation equations with respect to the radial variable. We show that for radial geodesics with $L=0$ this equations are integrated in quadratures for any spacetime dimension (\ref{genradsol}) and (\ref{genangsol}).

Next, we concentrate on metrics of a particular kind. First the Schwarzschild--Tangherlini metric defined in (\ref{ShTg}). It allows to express solutions in elementary functions in five-dimen\-sional spacetime (\ref{elsol5r}) for radial geodesic deviation vector component and (\ref{elsol5a}) for angular ones, while in six- and seven-dimensional space solutions are expressed in Weierstrass elliptic integrals (\ref{elsol6r}), (\ref{elsol6a}) and Legendre elliptic integrals (\ref{elsol7r}) and (\ref{elsol7a}) respectively. Thus, we have presented new solutions of the geodesic deviation equation at in multidimensional Schwarz\-schild spacetimes $D=5,6,7$ to the previously known solution in the 4-dimensional case in Ref.~\cite{SSSS}. In the case of higher dimensions, we expanded the solutions expressed in quadratures into power series. As expected, at spatial infinity, all components of the geodesic deviation vector grow linearly (\ref{eq36}) and (\ref{eq37}), regardless of the dimension $D$, this is due to the fact that the Schwarzschild-Tangherlini metric is asymptotically flat. On the contrary, in the vicinity of a physical singularity, there is a dependence of geodesic deviation vector behavior. Thus for $r\to0$ radial component $\xi^r\propto \frac{1}{\sqrt{r^{D-3}}}$, and angular components $\xi^a\propto r,\;a=\theta_1,..,\theta_{D-2}$. That is, as we approach the physical singularity, $\xi^r$ grows indefinitely, and all $\xi^a$ shrinks to zero. Therefore, we can conclude that test bodies moving in a gravitational field will experience stretching along the radial direction in higher-dimensional spaces, in transverse angular directions, tidal compression does not depend on the dimension of space, but the number of these components is determined by $D$, therefore, the number of direction along which the test body comprises.

Next we concentrate on Reissner--Nordstr{\"{o}}m--Tangherlini metric defined by (\ref{RNTg}). Explicit solution in 4-dimensional spacetime was described in \cite{TFrn}. We have succeeded in obtaining solutions of the deviation equation in five-dimensional space in terms of Legendre elliptic integrals (\ref{elsolch5r}) and (\ref{elsolch5a}). For dimensions greater than five, we present the solutions as power series. Thus, at spatial infinity, we once again obtain a linear growth of all deviation vector components (\ref{48}) and (\ref{49}), because the presence of an electric charge has not change the asymptotic flatness of spacetime. But in the vicinity of a physical singularity $r=0$ geodesic deviation vector behavior depends on spacetime dimension. So according to (\ref{50}) radial component $\xi^r\propto\frac{1}{r^{D-3}}$, and angular components according to (\ref{51}) $\xi^a\propto r,\;a=\theta_1,..,\theta_{D-2}$. This allows us to notice that in the leading order electrical charge presence changed the intensity of the longitudinal tension, but did not change the strength of the transverse compression.

In Subsec.~(\ref{EQfucl}), we generalize the solutions found in Ref.~\cite{NRtf} by solving the geodesic deviation equations (\ref{FuEQ}) in the presence of angular momentum $L\ne0$ using the Frobenius method for equations of the Fuchs class, since these equations have no solutions in quadratures. This made it possible to find the behavior of solutions in the vicinity of the origin $r=0$ in the leading order, taking into account the difference between the polar and azimuthal components in the presence of nonzero angular momentum. So in the Schwarzschild--Tangherlini metric in according to (\ref{58b}), (\ref{62b}), (\ref{66b}) it was found $\xi^r\propto\frac{1}{\sqrt{r^{D-3}}},\;\xi^a\propto r,\;\xi^{\theta_{D-2}}\propto const,\; a=\theta_1,..,\theta_{D-3}$. That is, in the leading order, the behavior of only the azimuthal $\xi^{\theta_{D-2}}$ component of the geodesic deviation vector has changed, while the behavior of the radial $\xi^r$ and all polar $\xi^a,\;a=\theta_1,..,\theta_{D-3}$ components has not changed. If a black hole has an electric charge in accordance to (\ref{71b}), (\ref{75b}), (\ref{79b}) it was found $\xi^r\propto\frac{1}{r^{D-3}},\;\xi^a\propto r,\;\xi^{\theta_{D-2}}\propto const,\; a=\theta_1,..,\theta_{D-3}$. Thus, in both considered Tangerlini metrics, we have found that the angular geodesic deviation vector components are independent of the spacetime dimension in the leading order in the vicinity of the black hole singularity. And only the radial components grow indefinitely as the radial variable $r$ decreases. This is in complete agreement with the conclusions of the work \cite{NRtf}, where for the first time local solutions of Eqs.~(\ref{FuEQ}) depending on L were considered only at $D=4$. Therefore, a test body freely falling along a non-radial geodesic undergoes unlimited stretching along the radial direction, and compression along all transverse directions, except for the azimuthal one. In considered case, the intensity of radial stretching in the leading order depends on the presence of a black hole charge and spacetime dimension, while the intensity of transverse compression does not depend on the presence of a charge or on spacetime dimension.

Tidal forces are an interesting and not yet fully understood phenomenon of gravitational interaction. There are a large number of works devoted to the study of tidal forces near black holes or separately to the geometric properties of the geodesic deviation equation. However, there is still a huge number of unsolved problems in the study of tidal forces. For example, no one has yet considered the tidal effects created by black rings. It is also worth noting that non-linear tidal effects remain unexplored, which can arise if we consider the generalized Jacobi equation \cite{NGGDE}, which may contain second or more powers of $\xi^{\mu}$ unlike the linear Eq.~(\ref{deq}).

\section*{Acknowledgement}
We would like to express our gratitude to Yuri Viktorovich Pavlov for meaningful discussions and useful advice.
\appendix
\section{\label{FEQ}Fuchs' equations and Frobenius method}
Second order differential equation for a complex variable function $y(z)$
\begin{equation}\label{FE}
y''(z)+p(z)y'(z)+q(z)y(z)=0
\end{equation}
has a regular singular point at $z=z_0$, if the coefficients $p(z)$, $q(z)$ have pole at $z=z_0$ not higher than the first and second order, respectively. In other words, the coefficients are decomposed into Laurent series as follows
\begin{subequations}
\begin{equation}
p(z)=\sum_{k=-1}^{\infty}a_k(z-z_0)^{k},
\end{equation}
\begin{equation}
q(z)=\sum_{k=-2}^{\infty}b_k(z-z_0)^{k}.
\end{equation}
\end{subequations}
In the vicinity of a regular singular point $z=z_0$ two linearly independent solutions of Eq.~(\ref{FE}) can be found using the generalized series
\begin{subequations}\label{powsol}
\begin{equation}\label{fsff}
y_1(z)=(z-z_0)^{\zeta_1}\sum_{k=0}^{\infty}c_k(z-z_0)^k,
\end{equation}
\begin{equation}
y_2(z)=(z-z_0)^{\zeta_2}\sum_{k=0}^{\infty}d_k(z-z_0)^k,
\end{equation}
\end{subequations}
where $\zeta_1$ and $\zeta_2$ are roots of quadratic equation
\begin{equation}
\zeta(\zeta-1)+a_{-1}\zeta+b_{-2}=0.
\end{equation}
If difference $\zeta_1-\zeta_2$ is not integer, then both solutions of Eq.~(\ref{FE}) presented by power series (\ref{powsol}), but if difference $\zeta_1~-~\zeta_2$ is positive integer, then the first solution for higher $\zeta_1$ remains the same (\ref{fsff}), however the second linearly independent solution has another form
\begin{equation}
y_2=(z-z_0)^{\zeta_2}\sum_{k=0}^{\infty}d_k(z-z_0)^k+Ay_1(z)\ln(z-z_0).
\end{equation}
\section{\label{ElInt}Elliptic integrals}
The simplest indefinite elliptic integral is the expression
\begin{equation}\label{elin}
\int R\left(z,y(z)\right)dz,
\end{equation}
where $y(z)=\sqrt{a_0+a_1z+a_2z^2+a_3z^3+a_4z^4}$ and $R(z,y)$ is rational function of two variables. Any elliptic integral can be brought to the form
\begin{equation}\label{elin2}
\int R_1(z)dz+\int\frac{R_2(z)}{y(z)}dz,
\end{equation}
where $R_1(z)$ and $R_2(z)$ are rational functions of one variable. Linear fractional transformation
\begin{equation}\label{lft}
z=\frac{ax+b}{cx+d}
\end{equation}
can turn the second term of (\ref{elin2}) into
\begin{equation}\label{elin3}
\int\frac{\tilde{R}_2(x)}{\tilde{y}(x)}dx,
\end{equation}
where $\tilde{R}_2(x)$ is another rational function, and $\tilde{y}(x)$ is reduced to Weierstrass form
\begin{equation}
\tilde{y}(x)=\sqrt{4x^3-g_2x-g_3},
\end{equation}
or Legendre form
\begin{equation}
\tilde{y}(x)=\sqrt{\left(1-x^2\right)\left(1-k^2x^2\right)}.
\end{equation}
As a result, expression (\ref{elin3}) can be represented as a sum of integrals of three kinds.
\begin{itemize}
\item In Weierstrass form integrals of three kinds are
\begin{subequations}\label{wi}
\begin{equation}
J_0(x|g_2,g_3)=\int\frac{dx}{\sqrt{4x^3-g_2x-g_3}},
\end{equation}
\begin{equation}
J_1(x|g_2,g_3)=\int\frac{xdx}{\sqrt{4x^3-g_2x-g_3}},
\end{equation}
\begin{equation}
H(x|g_2,g_3)=\int\frac{dx}{(x-c)\sqrt{4x^3-g_2x-g_3}}.
\end{equation}
\end{subequations}
\item In Legendre form integrals of three kinds are
\begin{subequations}\label{leg}
\begin{equation}
F(x,k)=\int\frac{dx}{\sqrt{\left(1-x^2\right)\left(1-k^2x^2\right)}},
\end{equation}
\begin{equation}
E(x,k)=\int\sqrt{\frac{1-k^2x^2}{1-x^2}}dx,
\end{equation}
\begin{equation}
\Pi(x,k,c)=\int\frac{dx}{\left(1-\frac{x^2}{c^2}\right)\sqrt{\left(1-x^2\right)\left(1-k^2x^2\right)}}.
\end{equation}
\end{subequations}
\end{itemize}
Where $c$ is some pole of the function $\tilde{R}_2(x)$, which can be complex.

\end{document}